\documentclass[%
reprint,
 amsmath,amssymb,
aps,
]{revtex4-2}

\usepackage{graphicx}
\usepackage{dcolumn}
\usepackage{bm}
\usepackage{amsmath, amssymb}
\usepackage[dvipsnames]{xcolor}

\begin{document}

\preprint{APS/123-QED}

\title{Measuring Optical Activity with Unpolarised Light: Ghost Polarimetry}

\author{S. Restuccia} 
\affiliation{School of Physics and Astronomy, University of Glasgow, Glasgow G12 8QQ, United Kingdom}
\author{G. M. Gibson}%
\affiliation{School of Physics and Astronomy, University of Glasgow, Glasgow G12 8QQ, United Kingdom}
\author{L. Cronin}%
\affiliation{WestCHEM, Department of Chemistry, University of Glasgow, Glasgow G12 8QQ, United Kingdom}
\author{M. J. Padgett}%
\affiliation{School of Physics and Astronomy, University of Glasgow, Glasgow G12 8QQ, United Kingdom}

\date{\today}

\begin{abstract}
Quantifying the optical chirality of a sample requires the precise measurement of the rotation of the plane of linear polarisation of the transmitted light. Central to this notion is that the sample needs to be exposed to light of a defined polarisation state. We show that by using a polarisation-entangled photon source we can measure optical activity whilst illuminating a sample with unpolarised light. This not only allows for low light measurement of optical activity but also allows for the analysis of samples that would otherwise be perturbed if subject to polarised light.
\end{abstract}

\maketitle

\section{\label{sec:level1}Introduction}
The optical activity of a sample is quantified by studying its interaction with a linearly polarised beam of light. In particular, when linearly polarised light passes through an optically active sample the plane of polarisation is rotated by an amount proportional to the degree of activity and is measured using a polarimeter. The technique is frequently used to measure the concentration or enantiomeric ratio of chiral molecules in solution \cite{Lough2002}.

In a traditional polarimetry detection scheme, the precision of the measurement increases with the intensity of the light used. This becomes a problem in situations where the intensity of the incident light might cause damage to the sample whose chirality is being probed, or indeed where the light itself may modify the chirality to be measured  \cite{Yoon2020}. A keen interest has therefore been growing in developing low light quantum detection schemes \cite{Tischler2016, Cimini2019}. It is also important to note that traditional polarimetry methods currently in use necessitate the probing light to be linearly polarised as the information on the optical activity is gained by observing the relative change in polarisation from before and after it interacts with the sample. In this work we propose an alternative detection scheme with a configuration similar to that which might be used for demonstrating Quantum correlations in the form of a Bell-inequality \cite{bell_einstein_1964, clauser_proposed_1969}. In these configurations, polarisation-correlated photons produced by parametric down conversion are employed to probe the chiral solution. We demonstrate that this system can be utilised for measuring the chirality in the low photon number regime and show that we are able to measure chirality using unpolarised light. 
\section{\label{sec:level1}Experimental Method } 
\begin{figure}[h]
  \centering
  \includegraphics[width=8.6 cm]{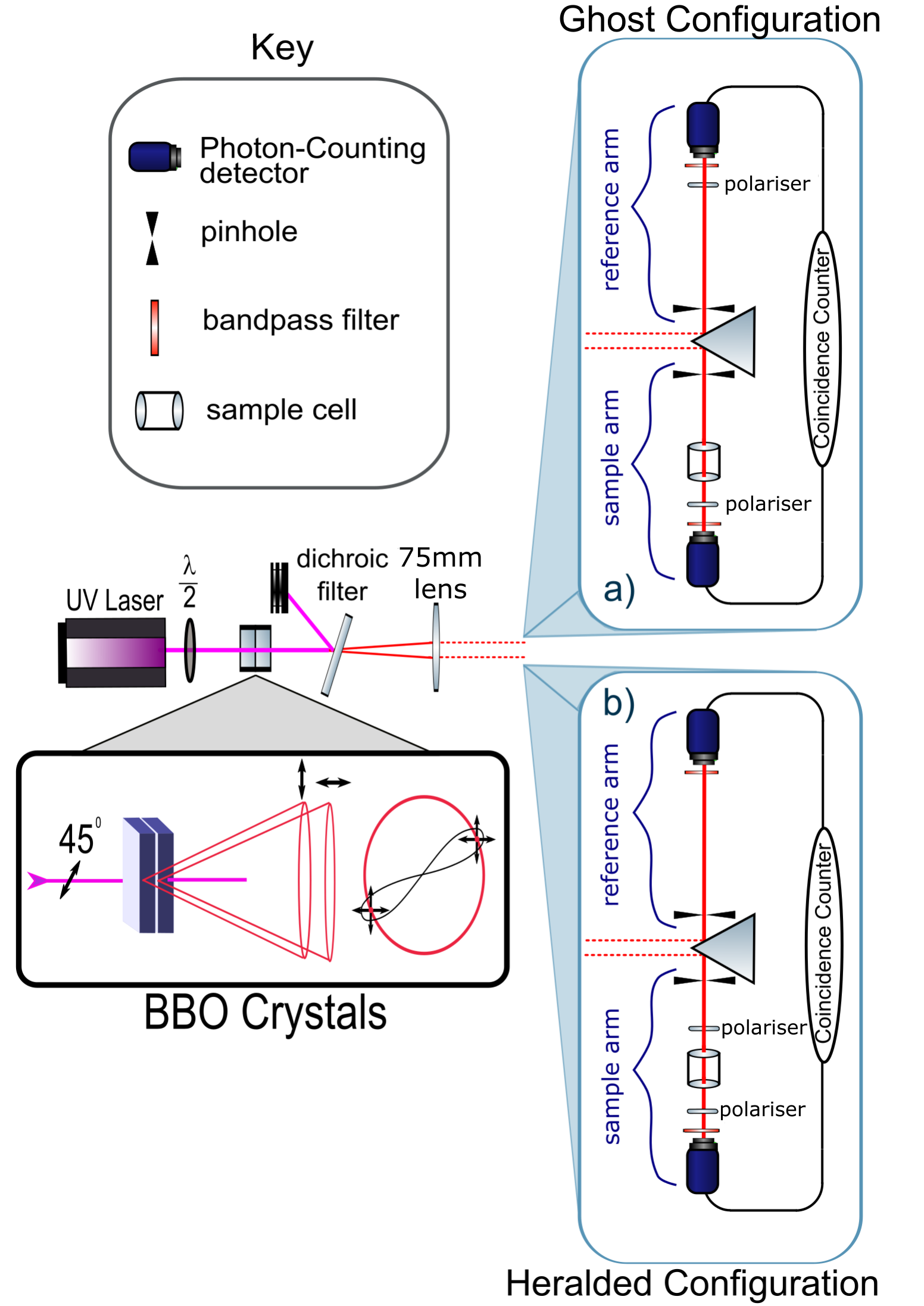}
\caption{Experimental set-up. A type-I two-crystal geometry SPDC source is employed to generate polarised entangled downconversion photons. The signal and idler components are then separated in the far field by a knife-edge prism into two arms each containing: a 1mm pinhole, a 10nm bandpass filter and a detector. When the system is in the Ghost configuration (a), two polarisers are employed one in each arm of the system. When the system is in the Heralded configuration (b), the two polarisers are placed in the same arm (sample arm), one at each side of the sample. For both configurations, coincidences are measured as a function of the relative polariser angles. When a sample cell containing the chiral solution is placed in the sample arm the optical activity can be measured.}
\label{fig:set-up}
\end{figure}
The quantum detection scheme proposed in this work uses polarisation entangled photons generated by a spontaneous parametric down-converted (SPDC) source of the type first described by Kwiat \textit{et al.} in their 1999 paper \cite{Kwiat99}. The down converted light is generated using a two-crystal geometry comprising of two type-I `sandwiched' 1mm crystals (with the optic axes perpendicular to one another), pumped by a 190mW, 355nm CW laser beam (as shown in figure \ref{fig:set-up}). A half waveplate is employed to orientate the pump beam at $45^{\circ}$  with respect to both the optical axes of the sandwiched down-converting crystals allowing for the the SPDC process to be equally likely to occur in either crystal i.e. generating a coherent superposition of vertically and horizontally polarised photons. The 710nm photon pairs generated through the SPDC process are then separated in the far field by a knife-edge prism into two arms of our experimental apparatus (sample arm and reference arm). Each beam then passes through a 1mm pinhole and interacts with a polariser before being collected for detection by photo-multiplier tube detectors. In particular we choose detectors with an effective diameter of 5 mm.

Although using a detector with a bigger aperture will inevitably lead to the collection of an increased number of background noise events, it allows for the measurement of photons with a wider range of acceptance angles making the experimental apparatus more resilient to small misalignments that can arise, for example, from swapping the sample cells in and out of the system. In order to minimise the number of detector events that correspond to background light a 10nm bandpass filter, centered at 710nm, is attached to the aperture of each detector and finally the two detectors are connected to a coincidence counter.

\begin{figure}[h]
  \centering
  \includegraphics[width=8.6 cm]{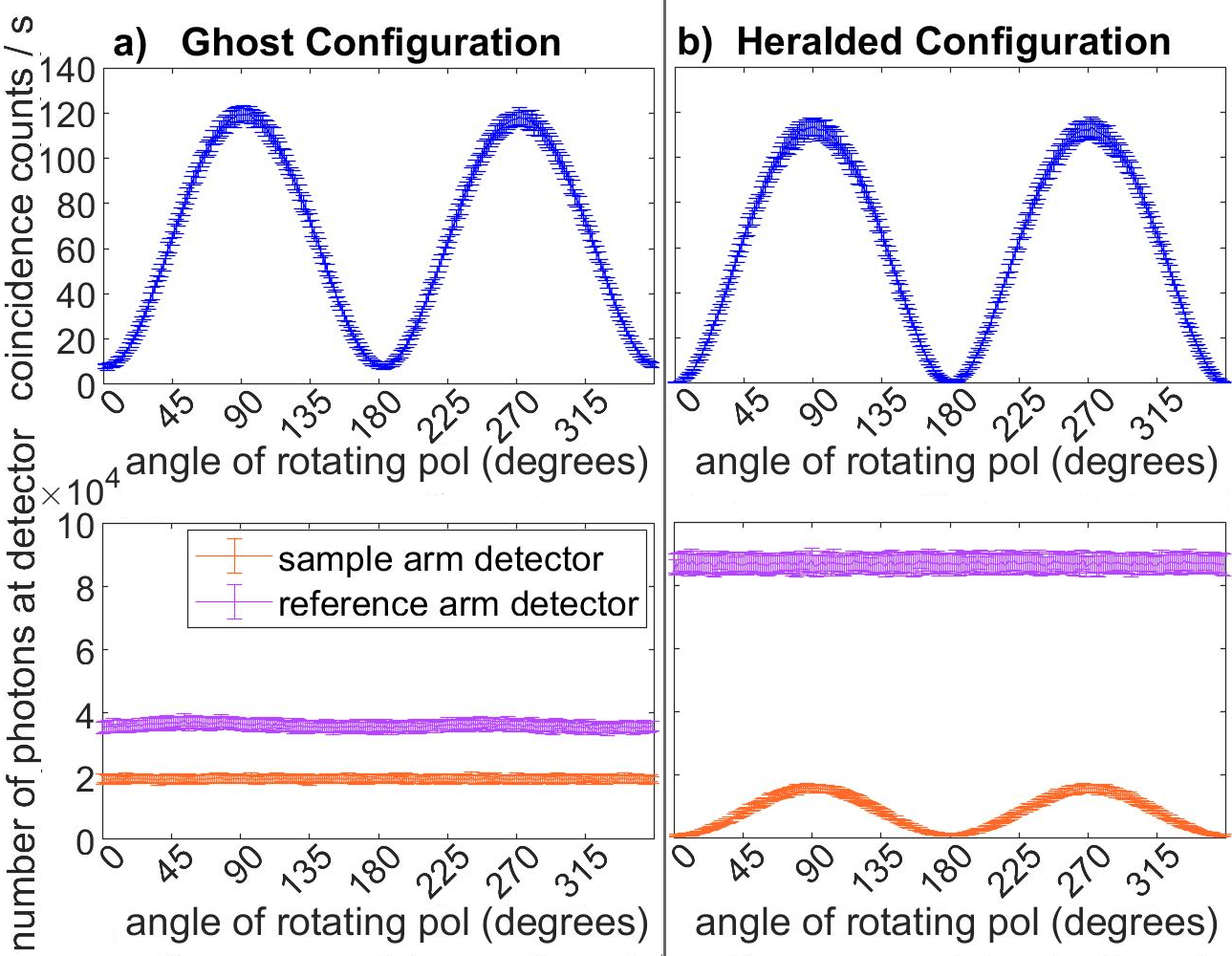}
\caption{Experimental results comparing the coincidence and photon counts obtained for both (a) Ghost configuration and (b) Heralded configuration. In the Ghost configuration a polariser is present in both arms and as the light is unpolarised the rotation of the polariser in the reference arm does not significantly effect the photon count rate, the sinusoidal form is only present in the coincidence counts. In the Heralded configuration no polariser is present in the reference arm (hence the photon count detected is higher) but both polarisers are present in the sample arm on either side of the cell. This will therefore produce a sinusoidal change in both the number of photons detected at the sample detector and the coincidence counts measured.}
\label{bellvsHerr}
\end{figure}

A coincidence counter is used to record both the single photon count rates for the pair of detectors and also to measure coincidences as a function of the polariser angles, thereby enabling for example a demonstration of the Bell-inequality. In the ghost configuration we typically recorded single photon count rates of S$\approx$19000 s$^{-1}$ (at the sample detector) and R$\approx$36000 s$^{-1}$ (at the reference detector) with corresponding coincidence count rate of $\approx$140 s$^{-1}$ (fig. \ref{bellvsHerr}a). With a gate time of $ \Delta t=1.523\times 10^{-9}$ s for the coincidence count rate this gives an accidental background count rate, $acc= S\times R \times \Delta t $, of $\approx$1 s$^{-1}$.

Drawing inspiration from works of Heralded- and Ghost-imaging \cite{Morris2015, MoreauReview}, we can see that there are two ways in which a down conversion source may be used to measure the optical chirality of a sample. In a heralded polarimetry configuration we insert both polarisers into the sample arm of the system, on either side of the sample (fig. \ref{fig:set-up}b). Rotating one polariser with respect to the other produces a sinusoidal variation in both the photon count rate measured by the detector in the sample arm and in the coincidence count between both detectors (fig. \ref{bellvsHerr}b). In particular, orientation of the transmitted polarisation was measured by fitting a sinusoid to 40 runs of the data to calculate both the mean phase (i.e. polarization orientation) and its standard deviation. It is important to note that compared to the photon count rate at the detector, in the case of the coincidence count the value measured is lower due to the finite heralding efficiency, and hence suffers from an increased shot noise, but has the net advantage of being largely background free. In particular it is well known that in the single mode definition for a coherent state, the shot noise level is defined by $<\sigma>/\sqrt{<n>}=1$ where $n$ is the number of counts and $<\sigma>$ is the standard deviation \cite{Brida2010,Kuzmich_1998}. In both the Heralded and Ghost configurations, the standard deviation increases with the average coincidence counts (fig. \ref{bellvsHerr}). In the particular case shown in fig. \ref{bellvsHerr} we choose to measure the coincidence count at each angle of rotation for 12s.

In the ghost polarimetry configuration (fig. \ref{fig:set-up}a) one polariser is inserted in each arm (as would be the case for a demonstration of the Bell inequality), but importantly the sample is placed before the polariser such that it is illuminated by the unpolarised down-converted light. In this case the single photon count rate is independent of the orientation of the polariser yet the coincident count rate retains its sinusoidal form (fig. \ref{bellvsHerr}a). In particular for the Ghost configuration we chose to rotate the polariser located in the reference arm of the set-up. As with the case of the Heralded configuration, the precision of the coincidence measurement in our ghost configuration is also limited by the shot noise on our detected signal.

In both the Heralded- and Ghost-configurations, the presence of a chirally active sample in the sample arm will result in a shift of the sinusoidal curves allowing for the direct measurement of the sample optical activity (as can be seen in section  \ref{sec:level4}).
 
One final question that arises is how we might expect the limiting precision of the measurement to scale with the number of coincidences measured.  This particular measurement problem can be considered from various perspectives. For a chiral sample, the incident and transmitted polarisation states are non-orthogonal and non-orthogonal state recognition is an established area \cite{Franke-Arnold2012}. Equivalently, the measurement of the sinusoidal variation in the count rate can be considered as a phase measurement problem. In the latter case the standard quantum limit for the estimation of phase, i.e. using  $n$ photons to probe the sample is given by $\Delta\phi=1/\sqrt{n}$ \cite{doi:10.1126/science.1104149}.

\begin{figure}[h]
  \centering
  \includegraphics[width=8.6 cm]{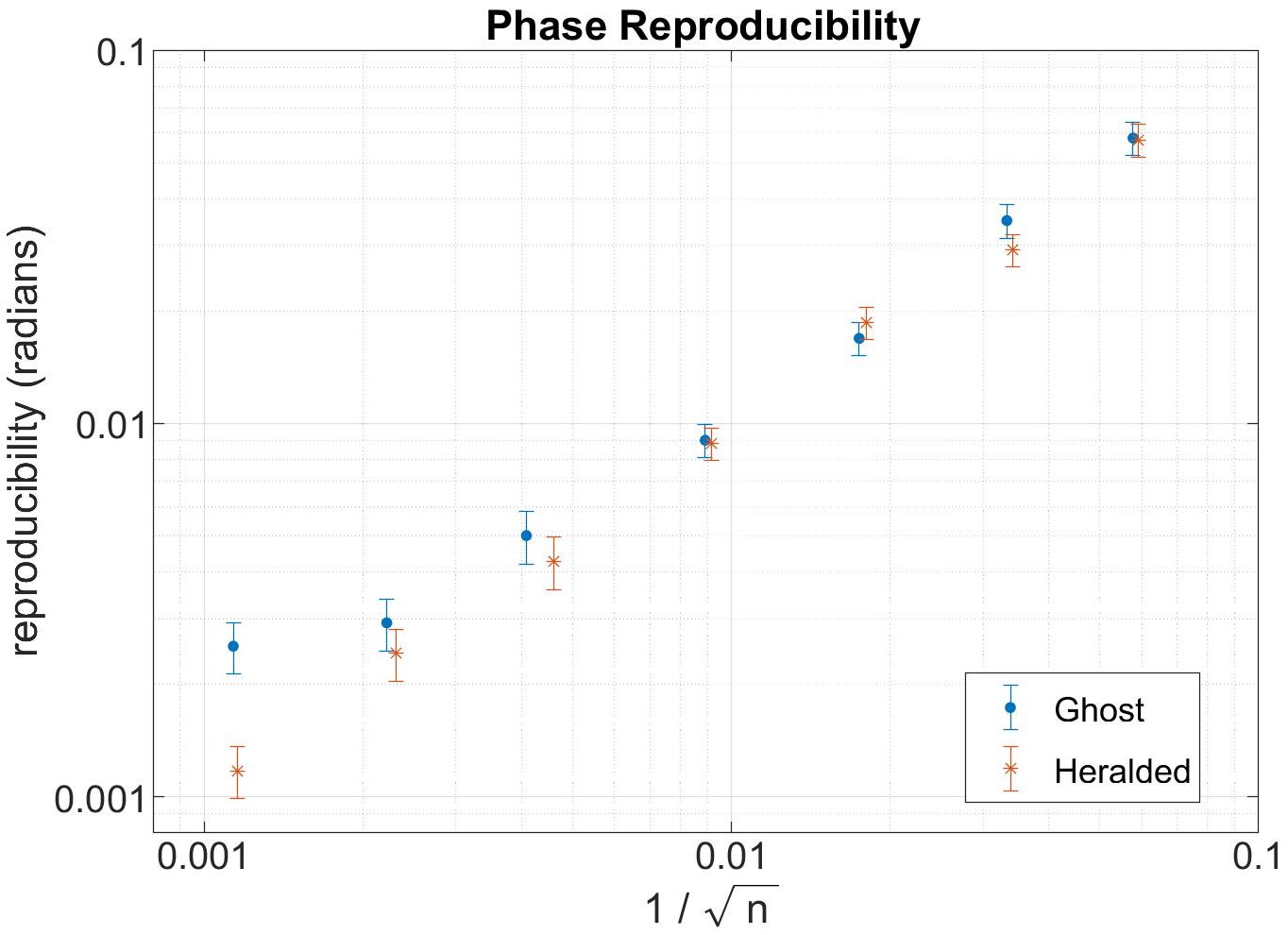}
\caption{For both the Heralded and Ghost configuration we measure the reproducibility of the phase measurement as a function of total coincidence counts calculated for a $2\pi$ rotation of the polarisation. The graph is represented in a log10 scale. As can be seen the graphs for both the Ghost (blue) and Heralded (yellow) configurations have a slope of approximately 1.}
\label{fig:errvsnumber}
\end{figure}

Figure \ref{fig:errvsnumber} shows the measured reproducibility of the polarisation orientation, as determined by the phase of the sinusoidal count rate as a function of the total number of coincident counts for a full rotation of the sample arm polariser.  The reproducibility of the curve is calculated by measuring the standard deviation of the phase after repeating the full sinusoidal measurement for a given number of times at different values of coincidence counts. The total number of coincidence counts used to measure the sinusoidal curve is regulated by changing the time integration used to calculate each data point in the curve. For the four  measurements in graph \ref{fig:errvsnumber} with the lowest coincidence counts ( $1/\sqrt{n}>0.01$) we repeated the experiment 100 times, while for the remaining data points we repeated the experiment 40 times. It must be noted that, because the polarisation vector is bi-directional, a rotation of the polarization state of $\theta$ advances the sinusoidal fringes by $2\theta$ so in calculating the phase of the sinusoidal curve we have to correct by a factor of two.  From figure \ref{fig:errvsnumber} we can see that for lower photon number the orientation uncertainty closely follows the anticipated relationship with photon number i.e. we are shot-noise limited. For very large numbers of photons the uncertainty is higher than expected. We believe this to be due to angular uncertainty introduced by the rotation stages used to rotate the polarisers and for which the uncertainty is  $\approx$ 1 mrad.

\section{\label{sec:level1}Experimental demonstration of the polarisation correlation}

To confirm the entangled nature of our photons and the fidelity of our experimental configuration we perform a Clauser-Horne-Shimony-Holt (CHSH) Bell inequality test \cite{bell_einstein_1964, clauser_proposed_1969}. We show that the state generated by our entanglement source violates a Bell inequality of the form
\begin{equation}
    |S|\leq2
\end{equation}
where
\begin{equation}
    S=|E(\theta_S,\theta_R)|+|E(\theta_S,\theta_R')|+|E(\theta_S',\theta_R)|+|E(\theta_S',\theta_R')|
\end{equation}
and

\begin{multline}
    \cr \textstyle E(\theta_S,\theta_R)= \\
    \scalebox{1.1}{
        $\frac{C(\theta_S,\theta_R)+C(\theta_S+\frac{\pi}{2},\theta_R+\frac{\pi}{2})- C(\theta_S+\frac{\pi}{2},\theta_R)-C(\theta_S,\theta_R+\frac{\pi}{2})}{C(\theta_S,\theta_R)+C(\theta_S+\frac{\pi}{2},\theta_R+\frac{\pi}{2})+ C(\theta_S+\frac{\pi}{2},\theta_R)+C(\theta_S,\theta_R+\frac{\pi}{2})}$}
\end{multline}

In Eq.3,  \(C(\theta_S,\theta_R)\) is the coincidence count measured when the polarisers in the sample and reference arm are rotated by $\theta_S$ and $\theta_R$ respectively. In particular, a CHSH inequality is maximally violated for the values: $\theta_S=22.5^{\circ}$, $\theta_R=0^{\circ}$, $\theta_S'=67.5^{\circ}$, $\theta_R'=45^{\circ}$ \cite{Poul2019}.

As can be deduced from Eq.2 and Eq.3 a full measurement of the CHSH test requires 16 measurements corresponding to different orientations of the polarisers. For our S calculation we therefore choose to calculate each of these measurements for a time integration of $\approx$3.5s. Repeating the whole measurement sequence 100 times we obtain a CHSH, S-value of $2.39+/- 0.07$. It should be noted that, providing the rotation of the polarization is accounted for, then inserting a chiral sample in our system does not destroy the polarisation entanglement of our photons. In particular, with a 1cm sample present in the system we still obtain a CHSH, S-value greater then 2 i.e. we measured an S of 2.46+/- 0.02. The greater precision in the measurement of S stems from two factors: as the presence of a chirally active sample result in a shift of the sinusoidal curves we choose to measure the full sinusoidal curve instead of only 16 measurements giving us a more precise measurment of the data point in eq.3 and we chose to only take 4 consecutive measurement sequence but integrate each measurement point for $\approx$12.2s. This corresponded to an $n=203000$ coincidence counts with which correspond to a phase reproducibility of $\approx$0.002rad.

\section{\label{sec:level4}Experimental measurements of optical chirality}

\begin{figure}[h]
  \centering
  \includegraphics[width=8.6 cm]{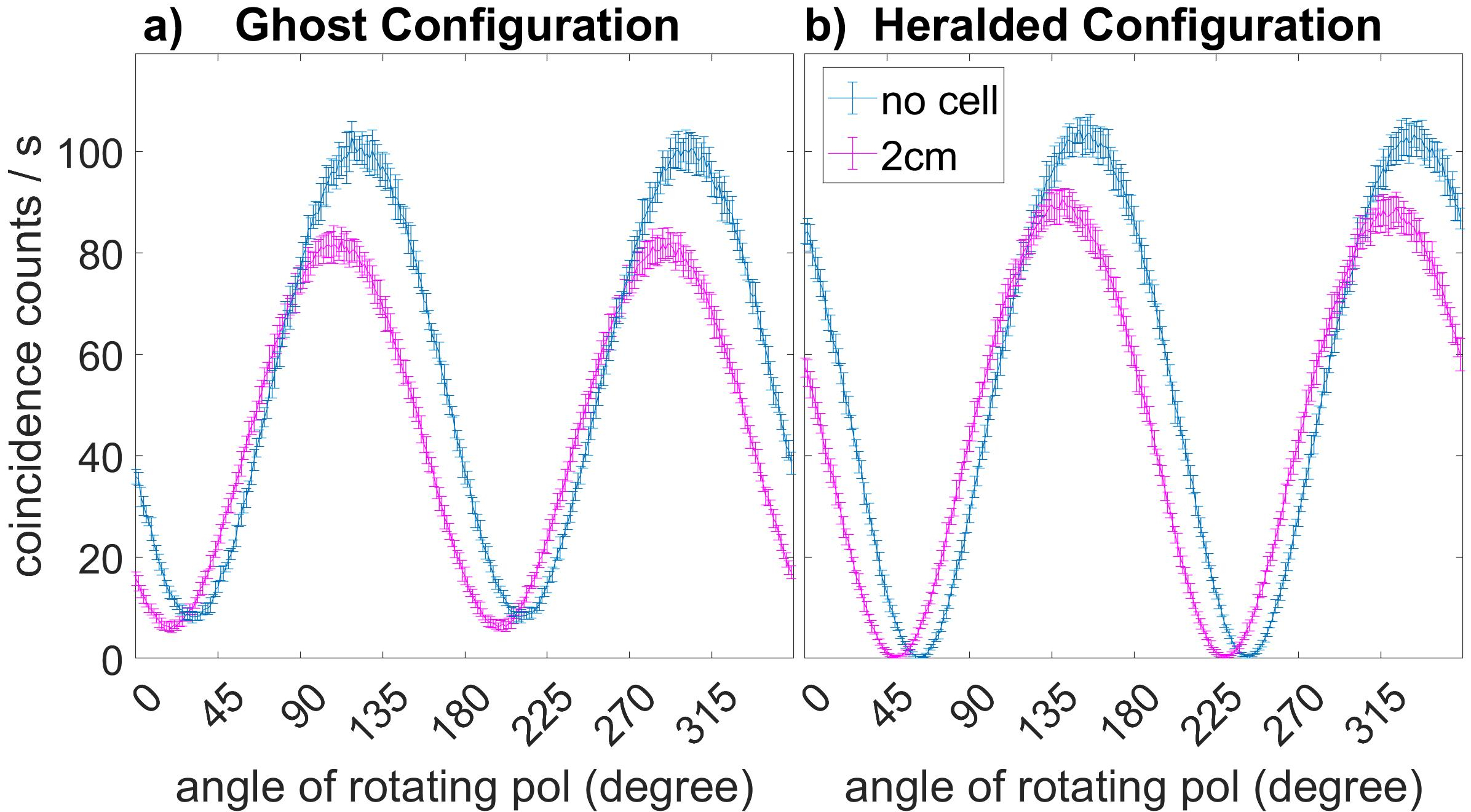}
\caption{Comparison between a coincidence measurement taken with no solution in the system and with 2cm of D-Limonene in the system in both the Ghost and Heralded configuration. The coincidence measurements are shown as a function of polarization angles. In both cases a shift of $~25$ degrees can be seen due to the chirality of the sample. Note that the coincidence counts with the sample present is lower due to the fact that inserting a glass cell in the system results in a loss of photons. All measurements where obtained with an acquisition time of $\approx$12.2s.}
\label{fig:graph}
\end{figure}

Finally, we demonstrate the ability of our Heralded and Ghost configuration in measuring chirality, in particular the optical activity of D-Limonene. The choice of D-Limonene as a sample is simply for convenience as it is an easily accessible highly chiral solution. As optical activity scales linearly with sample length, to accurately assess our measurement we repeated the experiment for sample cells of 1, 2 and 5cm in length. In the case of the Ghost configuration, the different cell lengths yielded a shift in the sinusoidal curve of 12.72$^{\circ}$+/-0.62$^{\circ}$, 25.1$^{\circ}$+/-0.77$^{\circ}$, 69.49$^{\circ}$+/-1.5$^{\circ}$ respectively. In the Heralded configuration we measured a shift of 12.14$^{\circ}$+/-0.56$^{\circ}$, 24.54$^{\circ}$+/-0.37$^{\circ}$ and 67.85$^{\circ}$+/-0.41$^{\circ}$ for the 1cm, 2cm and 5cm cells. It must be noted that the known optical rotation for a decimeter tube sample of D-Limonene at room temperature is 124\textdegree (i.e. 12.4$^{\circ}$ for a 1cm cell) \cite{lide_a._1995}. Figure \ref{fig:graph} shows the coincidence count rates as a function of the relative orientation of the polarisers in the Ghost and Heralded configurations with and without the presence of a 2cm chiral sample in the measurement arm. As expected the measurement of the optical activity of D-Limonene for both configurations yield a similar result as the optical activity of D-Limonene is independent of the polarisation of the light it is probed with i.e. using non polarised light should yield a similar result to using polarised light and hence we can benchmark the efficiency of our Ghost configuration measurement against the Heralded configuration. Finally, the error in our measurement is based on repeatability so it does not take in consideration the precision of the cell length and any error in the path the light takes while going through the cell. This second error is especially important as it is directly related to the length of the cell and how the cell is inserted in the system. Nevertheless, as can be easily seen by our experimental results, not only are our measurements with the Ghost and Heralded set-up consistent with each other and with the known D-Limonine optical rotation value, but as expected the values increase linearly with cell length. 
 
 \section{\label{sec:level1}Discussion and Conclusions}
 
 In this work we propose an alternative detection scheme for measuring the optical rotation of a sample. In particular we propose a quantum configuration known as the Bell-inequality and demonstrate that photons in our system are entangled both before and after interacting with a chiral solution (i.e. we can do a CHSH measurement and obtain an S-value S$>$2). More interestingly we are able to show that we can use the polarised entangled photons to probe a chiral solution with unpolarised light. Even more curiously, we can measure the optical activity of a sample with light that has not interacted with the sample (i.e. ghost polarimetry). 
 We can then benchmark our result against a more conventional measurement of chirality i.e. using polarised light by converting our system to a Heralded set-up. As expected the optical activity as measured in the heralded and ghost configurations are statistically consistent with each other as the sample used is not polarisation sensitive. We recognise that even at the shot-noise level this approach to measurement is slightly sub optimal since we are using only simple polarisers and only two detectors. This means that we are recording only 25$\%$ of the possible coincident counts. If we were instead to use polarising beam splitters with two detectors in each arm, this would improve the standard deviation of the measurements by a factor of 2 \cite{Tischler2016}. 
 Nevertheless, beyond our system  an interesting manifestation of the implications of quantum entanglement, our approach allows the measurement of chirality even of samples that would otherwise be perturbed if subject to polarised light.

\begin{acknowledgments}
MP and LC conceived the project. The experimental design was by SR and MP with additional input from GG. Results were analysed by SR and MP. SR wrote the manuscript with input from all other authors. We acknowledge financial support from the EPSRC (UK Grant No. EP/S019472/1) and  from The Leverhulme Trust.
\end{acknowledgments}

\bibliography{Bell.bib}

\end{document}